\documentclass[journal,comsoc]{IEEEtran}
\usepackage[T1]{fontenc}
\usepackage[nolist]{acronym}
\usepackage{cite}
\usepackage{amsmath,amssymb,amsfonts}
\usepackage{algorithmic}
\usepackage{graphicx}
\usepackage{textcomp}
\usepackage{xcolor}
\usepackage{url}
\makeatletter
\def\ps@IEEEtitlepagestyle{
  \def\@oddfoot{\mycopyrightnotice}
  \def\@evenfoot{}
}
\def\mycopyrightnotice{
  {\footnotesize
  \begin{minipage}{\textwidth}
  \centering
  Copyright~\copyright~2021 IEEE. Personal use of this material is permitted. Permission
from IEEE must be obtained for all other uses, in any current or future
media, including reprinting/republishing this material for advertising or
promotional purposes, creating new collective works, for resale or
redistribution to servers or lists, or reuse of any copyrighted
component of this work in other works
  \end{minipage}
  }
}

\begin{document}

\title{Enabling Bi-directional Haptic Control in\\Next Generation Communication Systems:\\ Research, Standards, and Vision}

\author{Chathura Sarathchandra, Kay Haensge, Sebastian Robitzsch, Mona Ghassemian, Ulises Olvera-Hernandez

\IEEEcompsocitemizethanks{
\IEEEcompsocthanksitem Chathura Sarathchandra, Sebastian Robitzsch, Mona Ghassemian are with InterDigital Europe, Ltd.
\IEEEcompsocthanksitem Kay Haensge is with InterDigital Germany, GmbH
\IEEEcompsocthanksitem Ulises Olvera-Hernandez is with InterDigital Canada, Lt\'{e}e, 

E-mail: \{firstname.lastname\}@interdigital.com}
}

\IEEEoverridecommandlockouts
\IEEEpubid{\makebox[\columnwidth]{978-1-5386-5541-2/18/\$31.00~\copyright2018 IEEE \hfill} \hspace{\columnsep}\makebox[\columnwidth]{ }}

\maketitle

\IEEEpubidadjcol

\begin{abstract}
Human sensing information such as audio (hearing) and visual (sight) or a combination thereof audiovisual are transferred over communication networks. Yet interacting sense of touch (haptic) and particularly the kinaesthetic (muscular movement) component has much stricter end-to-end latency communication requirements between tactile ends. 
The statements in this paper, to enable bi-directional haptic control, indeed follow the widely accepted understanding that edge computing is a key driver behind Tactile Internet aiming to bring control and user plane services closer to where they are needed. However, with an updated wider analysis of (pre)standardisation activities that are chartered around Tactile Internet, this paper highlights the technology gaps and recommends open research topics in this area.

\end{abstract}


\IEEEpeerreviewmaketitle
\acrodef{3GPP}{3rd Generation Partnership Project}
\acrodef{5GC}{5G Core}
\acrodef{5GS}{5G System}

\acrodef{AMF}{Access and Mobility Management Function}
\acrodef{AR}{Augmented Reality}

\acrodef{COINRG}{Computing In the Network Research Group}

\acrodef{DoF}{Degree of Freedom}

\acrodef{eRG}{evolved Residential Gateway}

\acrodef{gNB}{Next Generation NodeB}

\acrodef{FID}{Forwarding Identifier}
\acrodef{FQDN}{Fully Qualified Domain Name}

\acrodef{HIC}{Haptic Interpersonal Communication}

\acrodef{KPI}{Key Performance Indicator}

\acrodef{IIoT}{Industrial Internet of Things}
\acrodef{ITU}{International Telecommunication Union}
\acrodef{ITU-T}{\ac{ITU} Telecommunication Standardisation Sector}
\acrodef{ISOBMFF}{ISO Base Media File Format}

\acrodef{LAN}{Local Area Network}

\acrodef{MEC}{Multi-access Edge Computing}

\acrodef{NAS}{Non-Access Stratum}
\acrodef{NCS}{Networked Control Systems}
\acrodef{NF}{Network Function}
\acrodef{NFS}{Network Function Service}
\acrodef{NFV}{Network Function Virtualisation}
\acrodef{NFVRG}{\ac{NFV} Research Group}
\acrodef{NPN}{Non-public Network}

\acrodef{PIN}{Personal IoT networks}

\acrodef{QoE}{Quality of Experience}
\acrodef{QoS}{Quality of Service}

\acrodef{SBA}{Service-based Architecture}
\acrodef{SBI}{Service-based Interface}
\acrodef{SCP}{Service Communication Proxy}
\acrodef{SDN}{Software-defined Networking}
\acrodef{SDO}{Standard Development Organisation}
\acrodef{SFC}{Service Function Chaining}

\acrodef{TD}{Tactile Device}
\acrodef{TE}{Tactile Edge}
\acrodef{TI}{Tactile Internet}
\acrodef{TSN}{Time Sensitive Networking}
\acrodef{TSC}{Time Sensitive Communications}

\acrodef{UE}{User Equipment}
\acrodef{UC}{Use Case}
\acrodef{UDM}{Unified Data Management}
\acrodef{UDR}{User Data Repository}
\acrodef{UPF}{User Plane Function}
\acrodef{URLLC}{Ultra-Reliable Low-Latency Communication}

\acrodef{VNF}{Virtual Network Function}
\acrodef{VR}{Virtual Reality}

\acrodef{WG}{Working Group}
\acrodef{XR}{Extended Reality}
\section{Introduction}

\begin{figure*}[htbp]
    \centering
        \includegraphics[width=0.7\textwidth]{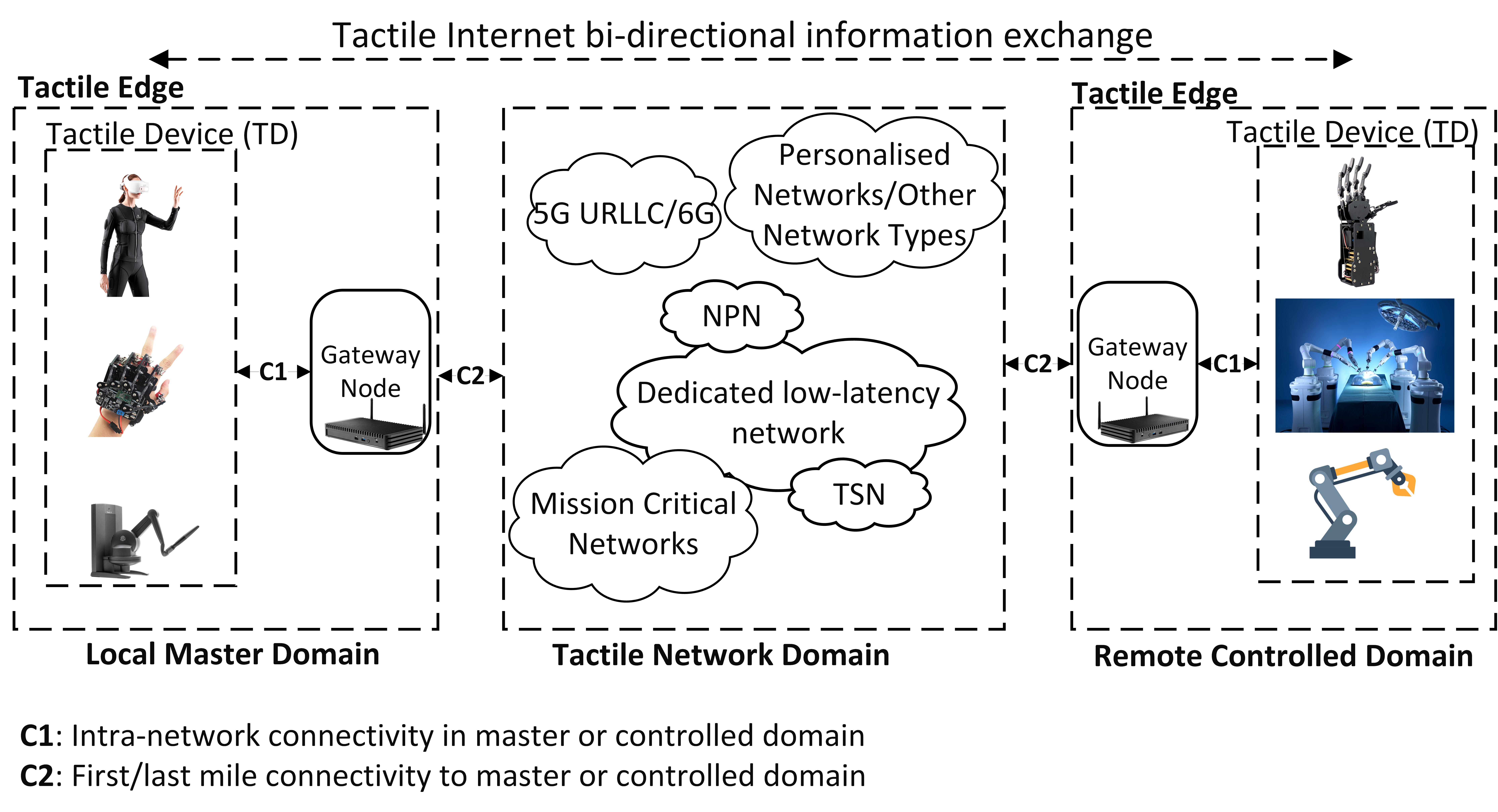}
        \caption{Overview of  Bi-directional Haptic Control \cite{8605315} \cite{9063407}}
        \label{fig:arch_figure}
\end{figure*}

\IEEEPARstart{H}{aptic} information in immersive communication enables mediated touch (kinaesthetic and/or tactile cues) over a computer/communication network to feel the presence of a remote user and to perform social interactions including handshake, pat, or hug. The application spectrum for haptic technology extends from social networking, gaming and entertainment to industry operation, training, and health care. The bi-directional haptic control system comprises a local master user, a remote controlled user, a remote user model at the local domain, and a local user model at the remote domain which can be split into three distinct domains as shown in as depicted in \figurename~\ref{fig:arch_figure} : the local master domain (including the operator and the haptic command interface), the tactile network domain (providing the medium for bilateral real feel communication between master and controlled domains), and the remote controlled domain (teleoperator  and the haptic feedback interface).  
The human models (remote participant or local user) can be either a physical entity (such as a social robot) or a virtual representation (such as a virtual reality avatar). 
Maintaining a human model for remote use involves the exchange of haptic data (position, velocity, interaction forces, etc.) and non-haptic data (gestures, head movements and posture, eye contact, facial expressions, user's emotion etc.).  
Typically, haptic information is composed of two distinct types of feedback: kinaesthetic feedback (providing information of force, torque, position, velocity, etc.) and tactile feedback (providing information of surface texture, friction, etc.). The former is perceived by the muscles, joints, and tendons of the body. The tactile feedback should not be confused with the \ac{TI} \cite{8605315} whereas the latter is consumed by the mechanoreceptors of the human skin. While the exchange of kinaesthetic information closes a global control loop with stringent latency constraints, this is typically not the case with the delivery of tactile impressions. In case of non-haptic control, the feedback is audio/visual and there is no notion of a closed control loop. 
In addition to enabling haptic/non-haptic control/data, the \ac{TI} aims to enable an overlay low latency network platform, providing  interoperability over different communication technologies, e.g., over 5G \ac{URLLC}, \ac{TSN}, etc.


The main contributions of this paper are in two folds: a) we provide a list of relevant standardisation activities addressing different aspects of bi-directional haptic communication and control, b) we highlight the technology gaps and open research topics to reach to the vision for realisation of the Tactile Internet. The rest of the paper is organised as following: Section \ref{sect:use-cases} discusses the \ac{TI} resource sensitive use-cases and requirements and presents early adaptors of the technology. Section~\ref{sect:related-work} presents related standards and bodies addressing different aspects of the \ac{TI} technology mainly on the architecture as well as data level. Existing technology enablers for a low latency communication platform are described in Section~\ref{sect:existing-tech} as required for the haptic data mode to be transmitted or perceived in immersive applications. Section~\ref{sect:research-outlook} highlights research and innovation opportunities to address the technology gaps to overcome the limitations as the result of distance projected by the speed of light when transmitting sense of touch over a distance. Finally, Section~\ref{sect:conclusion} concludes the paper and presents recommendations to the technology providers and standard organisations.
\section{Resource-Sensitive Use-Cases And Requirements }\label{sect:use-cases}

The TI will enrich the conventional audio/visual transport by including the ability of haptic control and provide the medium for transporting touch and actuation in real-time. The TI has a wide spectrum of use cases and applications ranging from use cases requiring \ac{URLLC} to ones with infrequent sampling of haptic data over less stringent networking modes.

In this paper, we focus on resource sensitive use cases, and define the operational states and technical requirements in which a Tactile Device (TD) should exist, over the course of its operation. Several \ac{TI} use cases are described and for each use case scenario, technical challenges, and advancements that \ac{TI} platform can provide are discussed. The use cases are surgical robotics (Tele-surgery), remote expert assistance in repairs (Smart factory) and multi-player interactive gaming and for \ac{VR}/\ac{AR} interaction, and human communications (Tele-presence).

The main application requirements for delivering a bi-directional closed haptic communication and control by the network are ultra-low latency, high availability, reliability, and security to enable the haptic data and feedback interactions. From a large range of applications discussed in literature, based on conducted interviews with different verticals, the early adopters of haptic technology are expected to be in the order of following: 
\begin{itemize}
    \item New generation of entertainment: Multiplayer games and entertainments where haptic wearables with sensory replacement, sensory augmentation enrich the interaction between the players in different locations,
    \item Cooperative manufacturing: Remote operation/mentoring of engineers to work from a safe distance to the hostile environment where multisensory (including \ac{XR} and haptic technologies) is required to enable the remote expert to guide at the precise location of the operation to the present worker in the factory,
     \item Interactive tele-care /tele-surgery: Tele-surgery and tele-mentoring for remote operation in rural areas where security, high reliability and extremely low latency communication is required to interact with haptic data/feedback.
\end{itemize}

The latency (end-to-end round-trip time) requirement is suggested to be in the milliseconds which represents the human reaction time. Yet for industry and robotic applications, the sensitivity of control circuits when controlling devices moving rapidly can require an end-to-end latency in sub-millisecond \cite{FG-NET2030-Sub-G1-usecases} per sensor/actuator. 

\section{Related Works and Standards}\label{sect:related-work}
This section provides an overview of standardisation bodies that explicitly work on \ac{TI} specifications or directly relate to \ac{TI} through architecture, use cases, focus groups, or study items.

\subsection{IEEE}
The IEEE \ac{TI} Standards \ac{WG}, designated the numbering IEEE 1918.1, undertakes pioneering work on the development of standards for the \ac{TI} as defined by the \ac{ITU} in August 2014. The IEEE \ac{TI} working groups initiated in 2016 \cite{8605315} aim to develop baseline standard addressing the network architecture as well as haptic codecs for the \ac{TI} providing a fast, reliable, secure, and available platform as part of the service requirements of the 5G and beyond.

The \ac{TI} haptic codec standardisation \ac{WG} provides codecs for the \ac{TI} enabling the interoperability of different haptic (kinaesthetic and tactile) input and output devices as required to achieve necessary market scale in the realisation of \ac{TI} technologies, devices, and applications.
Haptic codecs for the \ac{TI} address \ac{TI} application scenarios with human in the loop (e.g., teleoperation/remote touch applications) as well as machine remote control which define (perceptual) data reduction algorithms and schemes for closed loop (kinaesthetic information exchange) and open loop (tactile information exchange) communication. Furthermore, specific mechanisms and protocols for the exchange of the capabilities with respect to the workspace, the robot arm's number of \ac{DoF}, haptic signal amplitude range, temporal and spatial resolution of the haptic devices are included in the IEEE \ac{TI} standard scope. A fundamental challenge in context of the \acl{TI} is the development of a standard haptic codecs family, like the state-of-the-art audio (\ac{ITU-T} H.264) and video (ISO/IEC MPEG-4) codecs \cite{7593456}. Embracing both kinaesthetic as well as tactile information, such a codec family would be a key enabler for scalability at the network edge and universal uptake. Furthermore, it introduces a layered approach to haptic data (comprising multi-modal sensory information), which would be crucial for operation in typically challenging wireless environments.

Use case scenarios and requirements, technical assumptions, definitions, elements, functions, interfaces, and other related consideration are scoped in the architecture \ac{WG}. \ac{TI} architecture also includes the novel aspects and differentiating factors compared with, e.g., 5G \ac{URLLC} where it is noted that the \ac{TI} and associated requirements that the standard must serve to be likely operated as an overlay on other networks or combinations of networks.
\subsection{IETF}
Tactile Internet has been discussed in a number of groups within IETF primarily as a use case which demands for improving networking technologies towards satisfying its stringent resource requirements. 
Activities in \ac{NFVRG} suggests that a combination of radio access and core network components must be isolated into network slices for addressing specific requirements of emerging use cases, such as \ac{TI} services \cite{bernardos2018network}. \ac{COINRG} \cite{kunze2021coin} highlights requirements on in-network computing for providing real-time interactivity for immersive and mobile applications with tactile and time-sensitive data, which includes (but is not limited to), new internet architectures at the edge for improved performance, enabling joint collaboration, higher layer protocol optimisation to reduce latency and, enabling multi-stream, multi-device and multi-destination applications. 

A recent submission in DISPATCH working group \cite{muthusamy2021dispatch} introduces haptic data as a top-level haptic media type, and the recent acceptance of ‘haptics’ as a first-order media type in \ac{ISOBMFF}, making it historically the first attempt at registering haptics as a media format. Making haptics a top-level media type allows for the definition of data formats for haptic sub-modalities (e.g., kinaesthetic, vibrotactile) in a more streamlined manner. Given this development the authors make a case for haptic technology to be added to the list of top-level media types recognised by IETF.
\subsection{\ac{3GPP}}

From the outset, \ac{3GPP} \ac{5GS} were designed to provide highly reliable (e.g., \ac{URLLC}), service based, low latency communications (e.g., Edge Computing) and  enablers for Industrial Automation, e.g., \ac{TSC}, and Network Slicing. Although great progressed was achieved, many existing and new \ac{UC}s still remain to be addressed, e.g., \ac{UC}s with stringent requirements, as those needed to support  tactile  and  multi-modal  communication  services over  the  5G  system.  To address these challenges, \ac{3GPP} TR 22.847 intends  on  creating  a  gap analysis between new potential requirements and existing requirements and functionalities supported by \ac{3GPP}. Especially,for use cases that are immersive real-time experiences, including closed-loop feedback and control under varying \ac{DoF}s.

Requirements  for  cases  under  consideration  include (but  are  not  limited  to),  parallel transmission  of  multiple  modality representations associated with the same application. Also,  their  reliability,  availability,  security,  privacy,  charging, and  the  identification  of \ac{KPI} for specific use cases are considered. \ac{3GPP} TR 22.847 provides an example of new requirements to  further enhance \ac{3GPP} \ac{5GS} to meet the needs of demanding applications as those seen in the healthcare industry. 


\ac{3GPP} \ac{5GS} concepts, e.g., those explored in 3GPP TR 22.858 address deployments in residential environments, such as homes and small offices, includes both have wired and  wireless  converge. These concepts help us visualise where these systems may be deployed and where new requirements lay, including both \ac{3GPP} and Non-\ac{3GPP} Accesses. Services  deployed on fixed/wired networks differ from those deployed in mobile networks, as mobile  devices  tend  to  be  addressed  individually, while  in  fixed  networks,  devices  on  a  LAN  are  addressed through  a  gateway,  and  they  are  typically  not  known  to  the core  Network.  

In residential/small environments this gateway is referred to as \ac{eRG} and  they  are  considered  from  a  \ac{3GPP}  System  perspective as  \ac{UE}.  Nevertheless,  they  also  provide connectivity and \ac{QoS}  handling  to  other devices connected behind their realm, e.g., \ac{PIN}, which might correspond to sensors or actuators remotely controlled by e.g., a factory worker or a physician. It is thus quite  possible  that  in  an  End-to-End  scenario  where  a surgeon  wearing  a  tactile  glove,  which  is  a  wireless  device connected  through  an  \ac{eRG},  may  connect  to  actuators  in  a remote  location,  far  away  from  the  surgeon.  In  that  case, the  \ac{eRG}  at  each  end  of  the  connection  will  need to satisfy reliability and latency constrains, using technologies that may further extend the existing \ac{5GS}s

\subsection{ITU-T}
The Focus Group on Technologies for Network 2030, which concluded its work in Summer 2020, studied the capabilities of networks for the year 2030 and beyond. The group selected \ac{TI} as a representative use case for network 2030, among other use cases such as holographic type communications and Space-terrestrial integrated network \cite{FG-NET2030-Sub-G1-usecases}.


\section{Existing Technology Enablers and Testbeds}\label{sect:existing-tech}
This section presents the technology enablers for the Local Master, Remote Controlled and Tactile Network Domains as depicted in Fig 1. Furthermore, a number of related experimental testbeds are listed with a mapping to the discussed standards.
\subsection{Unified Data Link}
Any device requires a link local networking technology to communicate with another endpoint that has access to the same medium independent from its type, e.g., wired, wireless or optical. With the adoption of IEEE 802.3 - widely known as Ethernet - as the data link layer protocol for private and industry \acp{LAN} and eventually as the link layer protocol that runs the internet, many access technologies for devices followed suit such as WiFi, Bluetooth, Zigbee, LiFi and 3GPP's user plane with the release of 5G. Not only does such a unified access technology enable a rather homogeneous communication environment across a range of devices, it also lowers the complexity of supporting a range of data link layer in operating systems based on their physical layer realisation. More importantly, it allows \ac{TI} enabling technologies, such as \ac{TSN}, to work across all three \ac{TI} administrative domains (as illustrated in \figurename~\ref{fig:arch_figure}) for the delivery of packets in a guaranteed time, which is often referred to as "real time". 
\subsection{Advances in Communication Systems}
The Network Domain in \figurename~\ref{fig:arch_figure} illustrates a range of technology areas that are capable to deliver on the \ac{TI} requirements, as described in Section~\ref{sect:use-cases}. In the cellular telecommunication area, 5G has paved the way to support \ac{URLLC} use cases on the user plane and has even seen the adoption of cloud concepts in their system architecture aka \ac{SBA} in \ac{3GPP} TS 23.501, where the realisation of control plane functions follow cloud native concepts ensuring highly available, flexible and reliable services. Additionally, 5G systems come with a strong notion of \ac{QoS} enforcement procedures on the user plane combined with resource isolation procedures making it a significant technology enabler for interconnecting Tactile Edges (if \acp{TD} do not operate in the same Tactile Edge). Furthermore, the concept around \acp{NPN} can be seen as another technology enabler for \ac{TI} use cases where the 5G system is a fine tuned and focused solution for a specific use case without the need to support a wide range and diverse set of applications. 
\subsection{Edge Computing}
Moving services to the cloud has tremendously changed the landscape of service provisioning, reliability, and scalability. However, \acp{KPI} of sub-5ms latency and 1Gbps per user only allowed one logical move for services: if the network between client and servers cannot deliver those \acp{KPI} the service must come closer to the user. Combined with the advances of \ac{NFV} and private cloud solutions, edge computing is a key enabler in the context of \ac{TI} and offers a programmable and homogeneous framework to manage compute and virtual instances across these compute nodes. With the service orchestration frameworks that manage the lifecycle states of service instances across the edges, edge computing is a key technology enabler for achieving \acp{KPI} important for \ac{TI} services by bringing the service closer to the \ac{TD} or offering processing or storage intense service functionality to the edge.

\subsection{Terminals}\label{sect:terminals}
In the local master domain various sensors, actuators, display devices are used to provide a realistic haptic interaction with the remote devices in the controlled domain. The sensor components capture the tele-manipulation instructions (e.g., kinaesthetic) in the master domain, and the resulting changes in the control domain (e.g., haptic feedback) are shown in Figure \ref{fig:arch_figure}. Actuators, in both local master and remote control domains, execute the user’s tele-manipulation instructions. The number of independent coordinates used for providing the end user experience at the master domain (using Human System Interfaces), and for controlling the velocity, position, and the orientation of the controlled devices is defined by their \ac{DoF}. Today, there exist devices with different \ac{DoF} capabilities and are used match varying requirements of use cases \cite{8605315}. Advanced perceptual coding as well as resource management/task offloading techniques may be used for optimising communication and computing resources. 

\subsection{Testbeds, Demonstrations and Trials}
This section presents existing \ac{TI} testbeds, categorising ones that relate to \ac{TI} specific standards (IEEE 1918.1) in \ref{sect:testbed-std.oriented}, and ones that do not in \ref{sect:non-testbed-std.oriented}.
\subsubsection{Standards Oriented Testbeds}\label{sect:testbed-std.oriented}

An Extensible Testbed for \ac{TI} Communication (TIXT) \cite{9063407} implements a generic \ac{TI} architecture which is in line with the \ac{TI} architecture proposed in IEEE 1918.1 \cite{8605315} standards working group on \ac{TI}. By implementing a standards compliant testing environment it attempts to create a common ground for testing and evaluating TI research performed by various groups. TIXT’s implementation is highly extensible with its loosely coupled components using standard communication interfaces. \ac{TI} scenarios with a combination of simulators/emulators and real hardware can easily be incorporated into the same experiment.

The Haptic Communication Testbed at the Otto-von-Guericke university of Magdeburg (OVGU-HC) \cite{9217271} provides a data-driven experimentation platform. OVGU-HC has been designed targeting \ac{TI} experiments over small wireless networks as well as long-distance teleoperation scenarios over the internet. 
The platform provides a means for automating experiment scheduling and deployment through a generalised experiment description language. The authors demonstrate that the proposed platform is capable of implementing protocols and codecs presented in IEEE 1918.1 standards. 
\subsubsection{Non-Standards Oriented Testbeds}\label{sect:non-testbed-std.oriented}
A design approach has been presented \cite{8672612} for provisioning \ac{TI} services in \ac{VNF}-based \ac{MEC} systems. The authors develop, implement, and evaluate CALVIN, a low-latency management framework for distributed \ac{SFC}. A primary design choice aims at reducing computing latency by eliminating the context switching overhead due to shared virtual CPU between the kernel and user space, by implementing \acp{VNF} either purely in the kernel or user space, instead of utilising both together. Experimental results performed in a real testbed indicate that CALVIN can significantly reduce latency.

A \ac{SDN}- and \ac{NFV}-enabled experimental platform for 5G has been proposed for supporting \ac{TI} industrial applications \cite{8718538}. The testbed consists of three main layers (Backend, Field, and Network). The backend layer includes OpenStack based cloud resources where cloud applications are deployed. The network layer consists of SDN/NVF based industrial networks. The Network layer interconnects the compute nodes with the Field layer via virtualised \ac{IIoT} gateways and provides network slicing to support multiple applications. The virtualised \ac{IIoT} gateway interconnects smart sensors and smart light actuators with the backend cloud. It also enables to migrate \acp{VNF} down to the field, allowing for services with ultra-low latency requirements to be migrated to the edge for minimising latency.


TCPSbed \cite{8711100} presents a design of a testbed for Tactile Cyber Physical Systems (TCPS). Specifically, TCPS incorporates sensory feedback into CPS. Since, controlling latency and ensuring stability is crucial for TCPS applications, it allows isolation of latency of individual components of live applications, fine-grained characterisation of latency and control performance. Moreover, TCPSbed allows for both real and emulated networks and applications that mimic real-world embedded components for better emulating varying realistic conditions.

\section{Research and Innovation Opportunities}\label{sect:research-outlook}
There are a number of potential solutions for interacting haptic data within the latency constraints. In this section, we provide open areas for research and innovation from data, network, and terminal levels.
\subsection{Haptic Data Transmission and Perception}
There are a number of advanced techniques that can be exploited to support haptic data transmission and perception:
  
Artificial Intelligence (AI) and Machine Learning (ML) techniques allow creating a perception of the sense of touch within intelligent edge solutions for movement/haptic prediction to reduce latency. The multimodality of the situation can further support predicting the haptic information at the remote controlled domain either by learning the  movement models or predicting them using audio/visual information to reduce the perceived latency.
   
Secure and reliable transport solutions differ for haptic data. Various low overhead transport protocols are proposed and implemented by research communities with a few being standardised by IEEE, 3GPP, and IETF standard organisations or developed as a proprietary solution. Considering TI as an overlay network that can run different transport layer protocols and solutions in each part of the network, this can lead to interoperability issues and adding to the e2e delay.
    
Haptic data has been recently been accepted as a first-order media type in popular media formats such as MPEG ISO Base Media File Format \cite{muthusamy2021dispatch}. Efficient streaming techniques for transmitting haptic data along with any combination of existing (e.g., audio/visual) and new media types (e.g., ambience, emotion) must be further investigated (e.g., audio/haptic, audio/visual/haptic, audio/visual/haptic/emotion). Haptic sub-modality (e.g., kinaesthetic, touch) aware networking is another advanced open research topic. Haptic feedback can be different based on the application that it is being used for and the user interacting with the environment and the tactile device. Dynamic slice/edge relocation during a session based on user haptic experience feedback (e.g., telepresence of a family meeting may only use haptic feedback to communication emotion) or haptic interaction prediction based on other data modes (e.g. hug/handshake at the start and end) can improve the network resource utilisation. Additional signalling and functionalities are required to enable the utilisation of higher cost slice/edge resources only when needed during a session.

\subsection{Service-centric System Architecture}
The adoption of cloud native principles in 5G systems has a much greater impact on the overarching system design than simply turning monolithic network functions into microservices and orchestrate them into islands of microservice management systems inside the \ac{TI} network or Tactile Edge domain. One of the key design choices when realising an application in a cloud native fashion is the separation of routing, monitoring, analytics, and orchestration from the actual application whose main objective is to focus on processing incoming requests and returning a response accordingly. The naming ontology of services over the internet, i.e. \acp{FQDN}, is being used to allow the logical separation of functions which form the application. Thus, each client (\ac{5GC} consumer, \ac{UE}, \ac{TD}, gateway node or client process of a vertical application) is fully aware about the \ac{FQDN} of the next function (server) which can serve the request the client aims to issue. 

However, when distributing control or user plane (micro)services across a range of compute hosts (inside the \ac{TE} or inside the Tactile Network domain) and expecting to freely change their lifecycle or even orchestration state, challenges around dynamic and scalable name resolution and packet routing must be addressed through standardisation and adoption.

As the \ac{TE} can be equipped with a dedicated gateway node, as illustrated in \figurename~\ref{fig:arch_figure}, the network domain will see the gateway as the \ac{UE}. This results a hard boundary between \acp{TD} and the Tactile Network Domain which especially when it comes to network attachment, mobility and user authentication procedures (e.g., the \ac{NAS} protocol in 5G). For the realisation of a service-centric system architecture that sees everything as a service operating on top of a connectivity and routing layer, an evolution of \ac{NAS} and the likes are required to decouple their functionality from a static binding to a hardware component.

\subsection{Terminal Innovation}
New sensory and haptic functionalities required by applications of \ac{TI} further increases the demand on processing power required to execute the computation-intensive and low-latency tasks, while meeting stringent requirements of \ac{TI}. However, terminal/tactile devices (\ac{TD}) today have limited computing resources, and local processing of such tasks may not meet the latency requirements of \ac{TI}. Moreover, devices used by many \ac{TI} applications (e.g., robots, drones) contain only a limited supply of power. 
 
Offloading computation-intensive tasks from such resource scarce devices to be executed distributedly across resources over various networks (Fog, Edge, Cloud), towards minimising the response time and power consumption has recently gained popularity and has been identified as a crucial technique for improving \ac{TI} applications \cite{9203885}. \ac{TI} computation offloading frameworks must provide efficient function partitioning and offloading methods considering new device types (see section \ref{sect:terminals}) of \ac{TI} applications. Offloading algorithms must aim for minimising the total time spent on offloading. Moreover, improved resource allocation and management of resources at each constituent resource providers (e.g., Radio, Non-Radio, Computing) in the end-to-end system is paramount to reducing the total response time when offloading.

 


The incorporation of various human-system interfaces (\ac{TD} in Figure \ref{fig:arch_figure}) in \ac{TI} (with specific hardware functionalities) and the distributed execution of functions are key aspects of terminal evolution in \ac{TI}. The dynamic inter-connection of these distributed functions and \ac{TD}s can be seen as the composition of various hardware and software functionalities towards providing/improving the user's experience. The haptic and bidirectional communication brings consumption of experiences closer to humans (as opposed to experiences that are consumed today through mobile devices, for example). 
 
This increasingly distributed and human-centric experiences that are enabled by \ac{TI} require new and improved means for gathering and enforcing \ac{QoS} and \ac{QoE} metrics. Such requirements must be enforced throughout the distributed system, for improving not only the latency and reliability, but also the user’s haptic, visual, and audio experiences. However, for ensuring requirements in more dynamic scenarios where systems must adapt to satisfy/maintain requirements, dynamic methods for capturing both objective and subjective metrics of user experiences must be provided.

\section{Conclusion}\label{sect:conclusion}

Advancements in haptic technology, including high levels of co-presence in real time, which demand ultra-reliable low latency platforms over distance, offer high quality interpersonal communication experiences remotely. 
In this paper, we have provided an overview of the \ac{TI} research and developments in view of addressing further advancements required in the domain for realisation of the bi-directional haptic control capabilities in relevant industrial, entertainment and healthcare use cases. We have presented the related standardisation activities for haptic communication mainly by 3GPP, IEEE and IETF. A number of existing testbeds are discussed and categorised according to the standards used in the implementation. Considering the existing technology enablers, we highlight a number of open areas of research that a) are required for the interoperability between the discussed standards in order to realise a communication platform within millisecond and sub-milliseconds latency depending on the use case requirements, b) highlight advancements in other technologies (e.g., AI/ML) that can support improvement in haptic control scenarios, c) present terminal innovations that can evolve into immersive communications, including bi-directional haptic control.

While there are a number of activities in the standardisation organisations, \ac{TI} is not in the forefront in many \ac{SDO} groups and is exploited mostly as a use case. Further consistency and coordination across the various groups are required to establish and bring the various efforts by \acp{SDO} under one umbrella.


\ifCLASSOPTIONcaptionsoff
  \newpage
\fi

\bibliographystyle{./bibtex/IEEEtran}
\bibliography{./bibliography}

\begin{IEEEbiographynophoto}{Chathura Sarathchandra} is a researcher at InterDigital. He received his B.Sc. Hons. and Ph.D from the school of Computer Science \& Electronic Engineering, University of Essex, U.K., and he was awarded the UEssex scholarship for pursuing his Ph.D. Chathura has worked on various private, national, E.U. and international projects. He has held several IEEE \& ACM program committee memberships (e.g., ACM SIGCOMM, IEEE CloudCom, IFIP Networking). His current research interests fall within the general area of future computing technologies, including virtualization and mobile computing.
\end{IEEEbiographynophoto}

\begin{IEEEbiographynophoto}{Kay Haensge}
is a researcher at InterDigital Germany GmbH. He studied Information and Communication Technology at University of Applied Sciences in Leipzig and wrote his master's thesis about QoS-based WebRTC Service Delivery over a Next Generation Mobile Network (NGMN) Infrastructure. He worked as a researcher at the Deutsche Telekom AG to support the standardization group within the 3GPP\#SA2 TSG. After joining InterDigital, he continued working within EU-funded projects as well as support of standardisation activities within the scope of system architecture of 5G network elements. 
\end{IEEEbiographynophoto}

\begin{IEEEbiographynophoto}{Sebastian Robitzsch} is a Senior Staff Engineer at InterDigital and works on beyond 5G architecture innovations in the area of location-aware cloud-native orchestration, programmable systems combined with service routing capabilities. In the past he has been with Dublin City University, T-Systems, Fraunhofer FOKUS, and Nokia Research Centre. Sebastian received his Ph.D. from University College Dublin, Ireland, and an M.Sc. equivalent (Dipl.-Ing. (FH)) from the University of Applied Sciences Merseburg, Germany in 2008.
\end{IEEEbiographynophoto}


\begin{IEEEbiographynophoto}{Mona Ghassemian} is a senior R\&I manager at InterDigital.
She received her PhD from King’s College London (KCL) and continued at KCL as a research associate in E-Sense (FP6- EU project) for 2 years. She worked as a lecturer and senior lecturer at KCL, Uni of Greenwich and SBU, teaching courses and supervising at BSc, MSc and PhD levels for 12 years prior to her move to BT research in 2019. Mona is a senior member of IEEE, the IEEE UK \& Ireland section chair (2020-2021) and a member of IEEE Tactile Internet SA WG.
\end{IEEEbiographynophoto}

\begin{IEEEbiographynophoto}{Ulises Olvera-Hernandez} is a Senior Principal Engineer for InterDigital, responsible for developing 5G 3GPP System Architecture Evolution Standards. Prior to joining InterDigital, Ulises spend 15 year at Ericsson in Mexico, Canada, Sweden and Ireland developing Cellular Systems for International markets. Ulises holds a bachelor’s degree in Mechanical and Electrical Engineering from National Autonomous University of Mexico(UNAM).  Ulises has co-authored more than 150 granted patents, and he has more than 30 years of experience in the Research and Development of Cellular Systems, with emphasis in the development of the 3GPP 5G Next Generation Network and beyond.
\end{IEEEbiographynophoto}

\end{document}